\documentclass[amsmath,amssymb]{revtex4}
\usepackage{graphicx}
\usepackage{dcolumn}
\usepackage{bm}
\begin{document}
\title{The power spectrum of the 
circular noise}

\author{Daniel M\"uller}
\affiliation{
Instituto de F\'\i sica, UnB\\
Cxp 04455, 70919-970, Bras\'\i lia DF \\
Brazil} 
\email{muller@fis.unb.br}

\begin{abstract}
The circular noise is important in connection to Mach's principle, and also 
as a possible probe of the Unruh effect. In this letter the power 
spectrum of the detector following the Trocheries-Takeno motion in the 
Minkowski vacuum is analytically obtained in the form of an infinite series.
A mean distribution function and corresponding energy density are obtained 
for this particular detected noise. The analogous of a non constant 
temperature distribution is obtained. And in the end, a brief discussion about 
the equilibrium configuration is given.
\end{abstract}
\maketitle
\begin{center}
\end{center}

The non inertial vacuum, and the noise associated to it, has been extensively 
studied. It is called Fulling-Davies-Unruh effect, after the discoverers
\cite{1}. 

The more simple situation of proper constant acceleration, is already very 
well understood. For instance, a free scalar field can be quantized 
in Rindler coordinates. The vacuum obtained in these coordinates is unitary 
inequivalent to the Minkowski vacuum. 

The monopole type detector introduced by DeWitt \cite{1.2}, can give a meaning 
to these vacua. A detector at rest in Rindler coordinates is equivalent to an 
accelerated one in Minkowski space.  

As a consequence of Poincar\'e invariance, a detector in inertial motion in 
vacuum, does not get excited. This can also 
be understood in the detailed balance context of \cite{3}. 
The authors used the DDC \cite{4} formalism to obtain the Einstein 
coefficients for spontaneous excitation and emition. For non inertial motion 
the balance of the coefficients is upset, resulting in the Unruh effect.
The thermal character of the noise in the KMS sense, and the verification of 
other consistencies are reviewed in 
\cite{5}. In \cite{6} it is shown the connection between the absorptions of 
particles and the noise to which the detector is submitted.  

The rotational motion is different. Besides its own importance connected to 
the concept of inertia, this motion is of experimental relevance, 
because it provides a possible verification of the Unruh effect \cite{7}.    
The DDC formalism was also used in the context of rotation to obtain the 
spontaneous excitation of the detector, the circular Unruh effect \cite{10}.
In the very interesting work \cite{8}, De Lorenci and Svaiter develop the 
quantization  in Trocheries-Takeno \cite{85} coordinates, resulting in a non 
trivial vacuum. Davies {\it et al} also obtain a very interesting result \cite{9} 
concerning rotational motion. Since now, the 
De Lorenci-Svaiter-Trocheries-Takeno 
vacuum is the only one available for rotation. I do not attempt to enter into 
the vacuum question. 

In this letter, an analytical expression for the power spectrum in the form 
of an infinite series is obtained, a result which has not yet appeared 
elsewhere. In \cite{8} the transition rate obtained here, is left over as an 
integral. I suppose a detector in Minkowski vacuum, constrained to a 
Trocheries-Takeno type motion. It should be stressed that by following 
the procedure outlined in this letter, inside the light cylinder of \cite{9},
results in a transition rate similar to equation (\ref{eq3}) below.   

It is assumed a monopole type coupling of the detector and field,
\begin{equation}
L_I=-H_I=M\phi(x(\tau)), 
\end{equation}
where $M$ has energy levels. We require both the detector
and the field to obey Schr\"{o}dinger's unitary time evolution. Let us assume 
that in the very early past the detector is in the ground state with 
energy $E_0$, $|E_0\rangle$ and that the field is in the (Minkowski) vacuum 
state $|0_M\rangle$. As the detector moves, field and detector will undergo 
transitions to various states, in a trajectory dependent fashion \cite{2}. 
Since one does not care about the field's final state, the unknown final 
state of the field should be traced out from the transition probability for 
thetransitions  in question. This process yields, in first order 
perturbation  theory the following expression for the transition rate
\begin{eqnarray}
&&G^+(x_1,x_2)=\frac{1}{4\pi^2}\left(\frac{1}{-(t_1-t_2)+ 
(\vec{x_1}-\vec{x_2})^2}\right)\nonumber\\
&&G^+(x(s_1),x(s_2))=\langle 0_{M}|\phi^\dagger(x^{\mu}(s_1))
\phi(x^{\mu}(s_2))|0_{M} \rangle \nonumber\\
&&P(E)=C(E)\lim_{\tau_{0} \rightarrow \infty} \frac{1}{2
\tau_{0}}\int_{-\tau_{0}}^{+\tau_{0}} \int_{-\tau_{0}}^{+\tau_{0}}
e^{-iE(s_1-s_2)}G^+(x(s_1),x(s_2)) 
ds_1 ds_2,
\label{eq1}
\end{eqnarray}
where $E>0$ is the energy difference between the initial and final states of 
the detector and $G^+(x,x^\prime)$ is the positive frequency Wightman 
function.  

$C(E)$ depends on the internal constituency of the detector and a 
detailed discussion of it is given in \cite{3} and \cite{10}. While
the second term in (\ref{eq1}) corresponds to the noise this detector is 
submitted to, that's to say, on the way the field fluctuates as seen by the
observer in his trajectory $x^{\mu}(\tau)$. 

It is an easy exercise to show that a detector at rest in the 
Trocheries-Takeno coordinates $(dr^\prime=0,d\theta^\prime=0,dz^\prime=0)$
\begin{eqnarray}
&&t=t^\prime\cosh \Omega r^\prime-r^\prime \theta^\prime\sinh \Omega r^\prime 
\nonumber\\
&&r=r^\prime\nonumber\\
&&\theta=\theta^\prime\cosh\Omega r^\prime-\frac{t^\prime}{r^\prime}\sinh\Omega 
r^\prime\nonumber\\
&&z=z^\prime,
\end{eqnarray} 
according to (\ref{eq1}), has a transition rate given by
\begin{equation}
P(E) = -\frac{1}{16\pi^2\cosh^2\Omega r}\int_{-\infty}^{+\infty}
\frac{e^{-iEs}}{(s/2)^2 - [r/\cosh(\Omega r)]^2 
\sin[(s/2)\sinh(\Omega r)r^{-1}]^2}ds,
\end{equation}
where $s=t_2^\prime-t_1^\prime$ is the proper time between two points in the 
trajectory of the detector. After a change in the integration variable the 
above transition rate, is written as
\begin{equation}
P(E)=-\frac{\omega}{8\pi^2\gamma}
\int_{-\infty}^{\infty}\frac{\exp[-i(2E/(\gamma\omega))x]dx}
{(x)^2-v^2\sin(x)^2},
\label{eq2}
\end{equation}
where
\begin{eqnarray}
&&\omega=\frac{\tanh(\Omega r)}{r}, \omega\simeq\Omega,\;\;
r\rightarrow 0 \nonumber\\
&&v=\omega r,  v<1\nonumber\\
&&\gamma=\cosh(\Omega r)=\frac{1}{\sqrt{1-v^2}}. \nonumber
\end{eqnarray}
The fact that $v<1$, comes from the type of trajectory that the detector is 
following. In this sense, this trajectory is more physical, because the 
detector is not allowed to travel faster than the speed of light. 
In (\ref{eq1}), 
the $1/(2\tau_0)$ factor is associated to the adiabatic switching of the detector 
and the usual $-i\epsilon$ prescription is related to its size. As $v<1$, and 
the integral is over the real axis, the integrand in (\ref{eq2}) can be 
written as
\begin{equation}
P(E)=- \frac{\omega}{8 \pi^2 \gamma} \int_{-\infty}^{+\infty}
dx\frac{\exp[-i2(E/(\gamma\omega)) x]}{(x-i\epsilon)^2}
\sum_{n=0}^{\infty}\left(\frac{v\sin(x-i\epsilon)}{x-i\epsilon}\right)^{2n}.
\end{equation} 

After a binomial expansion, this last 
integral is evaluated using ordinary residue calculus 
\begin{equation}
P(E)=\frac{\omega}{2 \pi \gamma v}
\sum_{n=1}^{\infty}\sum_{k=1}^{\infty}
\frac{\left[v\left(-E/(\gamma\omega) +k\right)\right]^{2n+2k-1}
\Theta (-E +k\omega\gamma)}
{(-1)^{n-1}(2n+2k-1)(n-1)!(n+2k-1)!},
\label{eq3}
\end{equation}
where $\Theta(x)$ is the Heaviside step function, which  indicates 
that the rotational motion is the thermal reservoir.
The time scales of the detector $E,$ and the (proper) period of rotation of 
the detector  $\gamma T^\prime=2\pi/\omega$ are singled out in (\ref{eq3}), a 
property also shown in \cite{8} and \cite{9}. 

As a spectrum, (\ref{eq3}), can be divided by the energy $E$ and an {\it effective}
distribution function for the scalar particles can be obtained
\begin{equation}
n(E)=\frac{\omega}{2 \pi \gamma v E}
\sum_{n=1}^{\infty}\sum_{k=1}^{\infty}
\frac{\left[v\left(-E/(\gamma\omega) +k\right)\right]^{2n+2k-1}
\Theta (-E +k\omega\gamma)}
{(-1)^{n-1}(2n+2k-1)(n-1)!(n+2k-1)!}.
\label{dist}
\end{equation}
Effective here is in a certain sense, that the distribution function is dependent on 
the measuring apparatus. Other consequences connected to the motion, should be 
perceived by other types of detectors. Only the detectable properties are taken into 
account. This distribution function is understood as the mean number of particles per 
volume, with energy $E$ perceived by the detector. In the following, density will be 
assumed to be per unit volume. 
 
In this spirit, the total energy density of the scalar particles is
given by the following integral, which can be written in terms of the 
hypergeometric function
\begin{eqnarray}
&&E_t=\int_0^\infty En(E)dE=\frac{\omega}{2 \pi \gamma v}
\sum_{n=1}^{\infty}\sum_{k=1}^{\infty}\int^{k\omega\gamma}_0
dE \frac{\left[v\left(-E/(\gamma\omega) +k\right)\right]^{2n+2k-1}}
{(-1)^{n-1}(2n+2k-1)(n-1)!(n+2k-1)!}\nonumber\\
&&E_t=\frac{\omega^2}{2 \pi v}
\sum_{n=1}^{\infty}\sum_{k=1}^{\infty}\int_0^k dx
\frac{\left[v\left(-x +k\right)\right]^{2n+2k-1}}
{(-1)^{n-1}(2n+2k-1)(n-1)!(n+2k-1)!}\nonumber\\
&&E_t=\frac{1}{2 \pi r^2}\sum_{n=1}^{\infty}\sum_{k=1}^{\infty}\frac{(vk)
^{2(n+k)}}{(-1)^{n-1}(2n+2k)(2n+2k-1)(n-1)!(n+2k-1)!}\label{limite}\\
&&E_t=\frac{1}{2\pi r^2}\sum_{k=1}^{\infty} \frac{(\tanh(\Omega r)k)^{2k+2}
{}_{2}F_3(k+1/2,k+1;k+3/2,k+2,2k+1;-\tanh(\Omega r)^2k^2)}{(k+1/2)(k+1)(2k)!}\label{energia},
\end{eqnarray}
where $\Omega$ plays a role similar to the temperature, and there is an explicit 
dependence on the coordinate $r$. This corresponds to a non constant temperature 
distribution. For instance, in the above expression (\ref{energia}), at the origin when 
$r=0$, the energy density is $E_t=0$, which makes sense, because the motion of detector 
when $r=0$ is inertial. 

Next, the convergence of the infinite summation (\ref{limite}), is discussed. The second 
derivative with respect to $v$ of (\ref{limite}) is of the type
\begin{eqnarray} 
&& \sum_{k=0}^{\infty}(vk)^{2k+2}
  \sum_{n=0}^{\infty}\frac{(-v^2k^2)^n}{ v^2\Gamma(n+2k+1)n!}=
  \nonumber\\ 
&&=\sum_{k=0}^{\infty}(vk)^2(vk)^{2k}\frac{J_{2k}(2kv)}{v^2(kv)^{2k}}=
\nonumber\\
&&=\sum_{k=1}^{\infty}k^2J_{2k}(2kv)\nonumber\\
&&\rightarrow
  \sum_{k} \frac{k^{3/2}}{\sqrt{4\pi\sqrt{1-v^2}}}
\left(\frac{v e^{\sqrt{1-v^2}}}{1+\sqrt{1-v^2}} \right)^{2k}, \label{converg}
\end{eqnarray}
the last expression is valid asymptotically when $k\rightarrow\infty$ and $0<v<1$ 
\cite{WW}. In 
this asymptotic spirit, the summation (\ref{converg}) can be replaced by the integral 
over $k$, from $0$ to $\infty$, which results in
\[
g(v)=\frac{3\sqrt{\pi}}{4}\left|
\ln\left(\frac{v e^{\sqrt{1-v^2}}}{1+\sqrt{1-v^2}} \right)\right|^{-5/2}.
\]
The original summation (\ref{limite}) is obtained by integrating the last equation 
two times in $v$
\[
\int_0^v dv^\prime \int_0^{v^\prime} dv^{\prime \prime}g(v^{\prime\prime}), 
\]
which is plotted numerically in the following \\ \\
\begin{center}
\includegraphics[width=6cm,height=6cm]{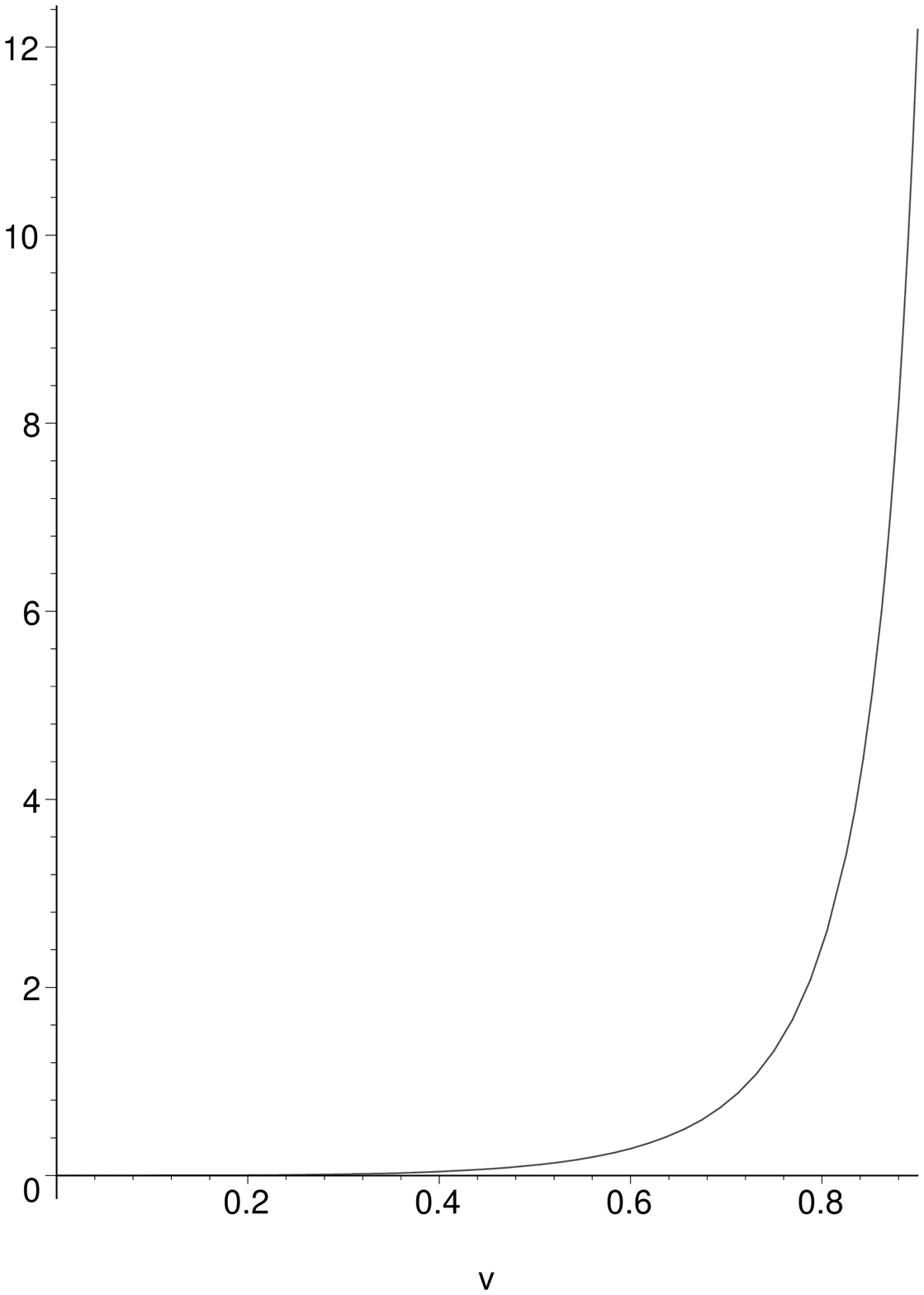}
\end{center}
The above graphic indicates, in a indirect way, that the summation given in (\ref{limite}) 
and (\ref{energia}) converges, except when $v\rightarrow 1$. This divergence of the 
total energy density of the scalar particles perceived by the detector, is expected when  
the detector is moving (almost) with the velocity of light, $v\rightarrow 1$. 

By summing the first few terms in (\ref{energia}), the following total energy density of 
scalar particles is obtained at a given $r$, for $\Omega=0.2$ and $\Omega=0.16$ \\ 
\begin{center} 
\includegraphics[width=6cm,height=6cm]{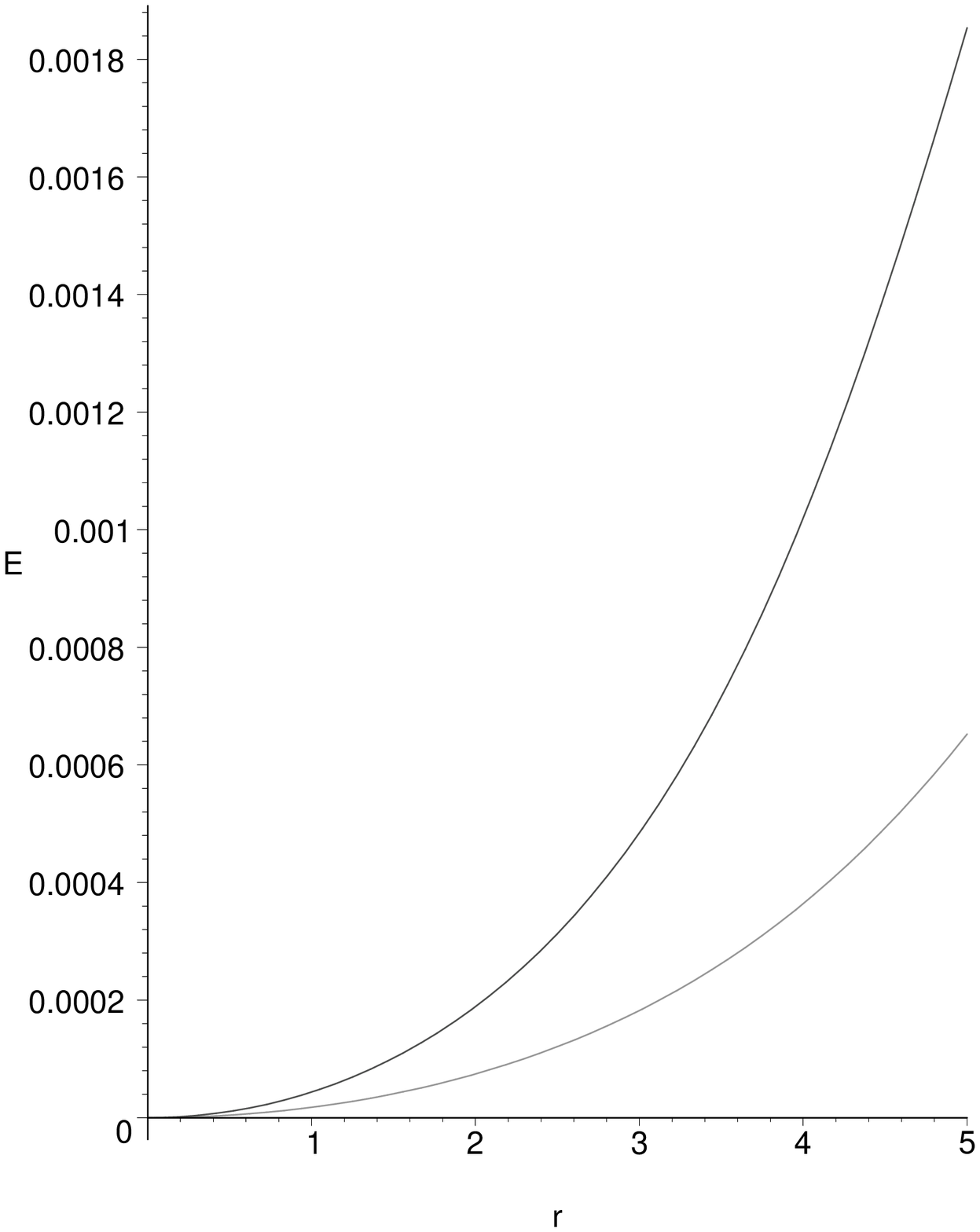}
\end{center} 
showing that for a same distance $r$ from the origin, there is an increase in the energy 
for larger values of $\Omega$.

It should be stressed that in spite that (\ref{energia}) describes a non constant 
temperature distribution, it is an  equilibrium configuration. 
This interesting equilibrium concept is well known and is expected if the space 
being considered is not homogeneous and isotropic as in Trocheries-Takeno case, see for 
example \cite{stephani}. The external condition that originates this motion is 
responsible for this apparent non equilibrium. The above construction avoids canonical 
quantization, and it could be usefull to a further understanding of external conditions. 

\section*{Ackowledgments}
The author acknowledges the brazilian agencies Funpe and Finatec for 
partial support, and an anonymous referee for improvements. 
\begin{flushleft}
{\bf Note added in proof}: It is a pleasure to thank Dr. Marcelo Schiffer which 
was my Msc. advisor.
\end{flushleft}

\end{document}